# Multi-element Persuasion in Social Media Health Communication: Synergistic and Trade-off Effects

Weifeng Zhang[a,b], Jipeng Tan[a,b], Mengye Yang[a,b], and Yong Min[a,b]*

[a] Computation Communication Research Center, Beijing Normal University, Zhuhai 519087, People's Republic of China.

[b] School of Journalism and Communication, Beijing Normal University, Beijing 100875, People's Republic of China.

*Corresponding author: Professor Yong Min, Computation Communication Research Center, Beijing Normal University, Zhuhai 519087, People's Republic of China. Email: myong@bnu.edu.cn

Notes on contributors:

Weifeng Zhang, a doctoral candidate at the Computation Communication Research Center, Beijing Normal University, focuses on computer communication, science communication, and health communication. [email: WeifengZhang993@163.com]| ORCID: 0009-0000-5170-8306

Jipeng Tan, a doctoral candidate at the Computation Communication Research Center, Beijing Normal University, focuses on computer communication. [email: 202431021021@mail.bnu.edu.cn]| ORCID: 0009-0000-5170-8306

Mengye Yang, a doctoral candidate at the Computation Communication Research Center, Beijing Normal University, focuses on computer communication, and social robot. [email: ymy6868@mail.bnu.edu.cn]

*Yong Min, a professor at the Computation Communication Research Center, Beijing Normal University, specializes in computer communication, media effects, and digital communication. He serves as the corresponding author. [email: myong@bnu.edu.cn] | ORCID: 0000-0002-9387-3921

# Multi-element Persuasion in Social Media Health Communication: Synergistic and Trade-off Effects

Health messages on social media are typically constructed through combinations of source cues, appeals, frames, and evidence, which jointly shape communication and persuasive effects. However, prior research has largely focused on single elements or simple pairwise interactions, offering insufficient insight into how multiple elements operate together in real-world digital environments. To address this gap, this study adopts a systems perspective to examine multi-element message combinations. Using 1.8 million health-related Weibo posts, we apply clustering analysis to identify recurring combinations and assess their relationships with communication effects. First, four recurring element combinations are identified: Institutional Authority, Narrative, Assertive Appeal, and Contextual Expression. These combinations function as core structures organized around two key elements. Second, stronger communication effects depend not only on core structures but also on peripheral elements aligned with these structures, with combinations of two to four peripheral elements generally showing greater advantages. Third, the optimal level of peripheral complexity varies with source influence, indicating that environmental factors condition the relationship between message combinations and communication effects. These findings show that communication and persuasive effects are shaped by synergies and trade-offs among multiple persuasive elements. Based on this, the study proposes a Core–Periphery–Environment framework to explain how message combinations generate communication effects with persuasive implications on social media. The study extends research from isolated elements to systems combinations and offers practical implications for health communication.

## Introduction

Health communication plays a central role in linking medical expertise to public well-being by shaping knowledge, attitudes, and behaviors (Scannell, 2021; Zou et al., 2021). With the rise of social media, this process has shifted into fast-paced environments where messages compete for attention (Chou et al., 2009; Moorhead et al., 2013). In such contexts, likes, comments, and shares serve as key indicators of communication effects, reflecting how audiences attend to, interpret, and respond to health information. (Alhabash & McAlister, 2015; Choi et al., 2021; Samuel, Kuijpers & Bleakley, 2024).

Health messages are rarely constructed around a single persuasive cue. Instead, they are composed of multiple elements, including source characteristics, emotional appeals, framing strategies, and types of evidence. These elements operate jointly rather than independently in shaping how messages are understood and evaluated. Prior research shows that affective cues tend to drive immediate reactions, whereas informational or utility-oriented content is more strongly associated with sharing (Kuo & Chen, 2025; Samuel, Kuijpers & Bleakley, 2024). This divergence suggests that communication effects depend on how these functions are coordinated within a message.

A substantial body of research has examined these elements individually, yet findings remain fragmented. Source credibility influences persuasion through expertise and trustworthiness (Petty & Cacioppo, 1984; De Meulenaer et al., 2017), emotional appeals shape cognitive and affective processing (Dillard & Nabi, 2006; Bilandzic et al., 2020), framing strategies guide interpretation through gain–loss distinctions (Kahneman & Tversky, 1979; Rothman & Salovey, 1997; Freling et al., 2020), and evidence type affects credibility and engagement, with statistical and narrative forms serving distinct functions (Green & Brock, 2000; Zebregs et al., 2015; Hinyard & Kreuter, 2007). Social media studies further show that content format, efficacy cues, and source attributes shape engagement patterns (Ta et al., 2022; Yap et al., 2019; Mello, 2023; Kuo & Chen, 2025; Jiang et al., 2024; Samuel, Kuijpers & Bleakley, 2024). However, these elements are better understood as interdependent components of message construction rather than independent predictors of persuasion (Limbu & McKinley, 2025).

This interdependence helps explain why prior findings are often inconsistent. The same element may produce different or even contradictory effects depending on its alignment with other elements. Existing research has examined interaction effects across evidence, source credibility, emotional appeals, and framing (Limbu & McKinley, 2025; Bilandzic et al., 2020; De Veirman et al., 2017). For example, evidence effectiveness depends on source credibility, with statistical evidence benefiting from credible sources and narrative evidence compensating for weaker source cues (Limbu & McKinley, 2025; Bilandzic et al., 2020). Interactions between appeals and framing further show that emotional effectiveness depends on contextual alignment, such as fear appeals in loss frames and hope appeals in gain frames (Dillard & Nabi, 2006; O'Keefe & Jensen, 2007). Narrative and statistical evidence also vary in effectiveness depending on message context (Hornikx, 2018; Braddock & Dillard, 2016).

Despite growing attention to interactions among message elements, most existing studies remain focused on single elements or simple pairwise combinations, often in controlled experimental settings. In real-world social media environments, however, messages typically integrate multiple elements simultaneously, forming structured combinations whose effects may reinforce or conflict depending on their coherence (Miller & Page, 2007; Bilandzic et al., 2020). Prior research identifies systematic interactions across source, appeal, framing, and evidence, particularly between evidence and credibility, emotional appeals and framing, and appeals and framing. Yet these interactions are usually examined separately, leaving many potential combinations unexplored. A systematic review further suggests that persuasive effects depend on how these elements are jointly configured, as their effects may reinforce or offset one another in shaping communication effects (Limbu & McKinley, 2025). This is consistent with prior research showing that evidence, appeals, and framing can produce synergistic or trade-off effects depending on alignment (Bilandzic et al., 2020; O'Keefe & Jensen, 2007; Hornikx, 2018; Braddock & Dillard, 2016). This combinational perspective implies that persuasion operates through coordination rather than accumulation: source provides credibility, appeals activate affective responses, framing guides interpretation, and evidence offers epistemic grounding (Mang, Fennis & Epstude, 2024; Xu et al., 2025; Choi et al., 2025). It also introduces variation in complexity, as additional elements raise the potential for both reinforcement and conflict, suggesting that persuasive outcomes may follow a non-linear rather than additive pattern.

In addition to message composition, communication and persuasive effects are shaped by contextual conditions such as source influence. On social media, follower size signals prominence and credibility, affecting both visibility and engagement (De Veirman et al., 2017; Liu & Zheng, 2024). Emerging evidence suggests that source influence moderates how message elements translate into audience responses, shaping the degree to which combinations of elements are required to achieve persuasive effects (Limbu & McKinley, 2025; Wang et al., 2023; Zannat et al., 2024).

Taken together, existing research highlights the importance of both multi-element message composition and contextual conditions but does not fully capture how multiple elements co-occur and operate as structured combinations in large-scale digital environments. This suggests that persuasion in social media contexts operates as a combinational process, where communication effects depend on the alignment among persuasive elements rather than their independent contributions.

To address this gap, this study adopts a systems perspective and data-driven approach using large-scale Weibo data. It treats health messages as interdependent combinations of elements rather than isolated cues, consistent with systems-based perspectives that emphasize interactions and emergent outcomes (Holland, 1992; Kitano, 2002). Rather than focusing on isolated elements, the study identifies recurring patterns of element combinations and examines how their internal coordination relates to different communication effect indicators under varying levels of source influence.

Accordingly, this study addresses the following research questions:

RQ1: What recurring patterns emerge from combinations of persuasive elements in social media health communication?

RQ2: How are synergies, trade-offs, and combinational complexity among persuasive elements associated with different communication effect indicators?

RQ3: How do the associations between persuasive element combinations and communication effects vary across different levels of source influence and combinational complexity?

## Methods
### Data Collection

This study uses Sina Weibo as the research context, one of the largest social media platforms in China with over 580 million monthly active users (Weibo, 2024). Its diverse user base and widespread use in public health communication make it suitable for large-scale health information analysis (Chou et al., 2018). Compared with highly visual platforms such as Instagram, Weibo is more text-centric and suitable for analyzing textual message composition. We focused on health-related content produced by verified accounts.

A total of 487 certified health science bloggers were identified based on platform verification and domain classifications on Weibo (Weibo, 2025). Using Python scripts to access the Weibo API, we retrieved 4,533,256 historical posts published by these accounts. To reflect original message construction, only first-level posts were retained, excluding reposts and forwarded content. Both standard posts and extended long-form posts (Changweibo) were included, with a customized script retrieving full text.

A multi-step preprocessing procedure was applied to improve data quality. Duplicate and non-original posts were removed, and a BERT-based classifier was used to retain health-related content. The final dataset consists of 1,813,438 valid posts, providing a robust basis for examining message composition and its associations with communication effects.

Although this study focuses on textual content, this choice is grounded in the central role of text in persuasive communication. Textual expression conveys key persuasive components, including scientific evidence, narrative structures, and emotional appeals (Semino et al., 2023). In health communication, framing, appeals, and evidence are typically operationalized through text (Ma & Nan, 2019; Wang et al., 2023; Zannat et al., 2024; De Meulenaer et al., 2017). This focus is appropriate for Weibo, where text remains central and enables scalable analysis of message composition.

### *Content Coding and Pattern Identification*

Message elements were coded using the SAFE framework (Source–Appeal–Frame–Evidence) (Limbu & McKinley, 2025). Each post was coded across four dimensions. Source was categorized as expert, official media, or no identifiable source (Vraga & Bode, 2017). Appeal included rational, emotional, value-based, and metaphor-based appeals (Dahlen et al., 2010; Rucker, Petty & Briñol, 2008; Hornik et al., 2016; Limbu & McKinley, 2025). Frame was coded as gain, loss, or no frame (Rothman & Salovey, 1997; Kang & Lee, 2018). Evidence was categorized as narrative, statistical, expert, causal, or no evidence (Allen & Preiss, 1997; Hoeken & Hustinx, 2009; Zillmann & Brosius, 2000). For each dimension, a "no" category captured absence.

Coding used a GPT-5-assisted approach (Whitty, 2025), with SAFE categories and conceptual definitions embedded in the prompts (Dunivin , 2025). These definitions, summarized in Table 1, served as an explicit codebook for AI-assisted deductive coding. Reliability was assessed using a stratified random sample of 1,000 posts coded by two trained graduate coders. Intercoder agreement reached acceptable levels, and agreement between human and GPT coding showed substantial reliability (Cohen's κ = 0.78; accuracy ≈ 83.5%).

Each post was then represented as a multi-label feature vector indicating the presence or absence of persuasive elements. A post could contain multiple features simultaneously, including a source type, one or more appeal types, a frame category, and one or more evidence forms. Clustering was performed on these coded vectors rather than raw text to identify recurring persuasive element combinations and co-occurrence patterns, instead of topic- or semantics-based clusters. HDBSCAN detected dense regions in this feature space without pre-specifying the number of clusters (Campello et al., 2013). The overall workflow is illustrated in Figure 1.

### *Measures of Communication Effects*

In social media research, communication effects are commonly measured using observable indicators such as likes, comments, and shares (Alhabash & McAlister, 2015; Choi et al., 2021; Wang, Lei & Xiao, 2026). Given the highly skewed distribution, all variables were log-transformed using $\log(x + 1)$ to reduce the influence of extreme values and enable more stable comparisons. In this study, likes, comments, and shares are treated as observable indicators of communication effects. They do not directly measure attitude or behavior change, but they provide indirect evidence of how persuasive message elements are taken up by audiences through attention, endorsement, interaction, and diffusion responses (Vraga & Bode, 2017; Lee, 2018; Alhabash & McAlister, 2015; Jiang et al., 2024).

### *Source Conditions*

Follower counts exhibit highly skewed, long-tail distributions, where a small proportion of accounts capture disproportionate attention (Zhu & Lerman, 2016). Rather than treating follower count as continuous, source visibility was operationalized using a distribution-based grouping approach. Accounts were divided into three tiers: top 10%, 50th to 90th percentile, and bottom 50%. This approach aligns with prior research conceptualizing influence in tiers (De Veirman et al., 2017; Lou & Yuan, 2019) and enables comparison across visibility conditions.

To validate this classification, we examined follower distributions across tiers. Results show clear separation, with mean follower counts increasing from bottom to top tiers (bottom: M = 256,227, SD = 88,007; middle: M = 1,047,240, SD = 524,313; top: M = 3,957,694, SD =

1,595,701). Within-group dispersion was limited compared to between-group differences. A one-way ANOVA confirmed significant differences across tiers, F = 764.83, p < .001, supporting the distinctiveness of the classification.

## Results
### *Recurring Message Patterns in Social Media Health Communication*

Using HDBSCAN, 21 clusters were identified from 1,813,438 health-related Weibo posts, with 26.3% classified as noise. Cluster sizes ranged from 284,927 to 19,193 posts (Figure 2 and Figure 3). To improve efficiency, the model was fitted on repeated subsamples (300,000 posts) and extended to the full dataset using approximate prediction.

To address RQ1, clusters were interpreted based on dominant SAFE feature combinations and grouped into four higher-order patterns, namely Institutional Authority, Narrative, Assertive Appeal, and Contextual Expression (

Table 2).

**(1) Institutional Authority Pattern**

This pattern is defined by the joint presence of authoritative sources and evidential support, including expert or official sources combined with expert or statistical evidence (over 360,000 posts).

Three subgroups were identified: (1) expert sources combined with value or utilitarian appeals and expert evidence (Clusters 0, 15, 16); (2) expert sources with evidence but limited appeals (Clusters 7, 14); and (3) official media combined with expert-related cues and utilitarian or value appeals (Clusters 13, 20). Overall, this combination is centered on source and evidence.

**(2) Narrative Pattern**

The Narrative pattern centers on narrative evidence presented through stories or experiences (approximately 160,000 posts).

Subgroups include practical narratives combining narrative evidence with value or utilitarian appeals (Clusters 1, 3, 9), Daily narratives with weak source and appeal cues (Clusters 4, 5), and authority-linked narratives combining narrative structure with expert signals (Cluster 6). Narrative evidence is the defining feature.

**(3) Assertive Appeal Pattern**

This pattern is characterized by the absence of explicit source and evidence, relying instead on direct appeals such as value, utilitarian, or fear-based messaging (approximately 500,000 posts).

Subgroups include practical assertive messages (Clusters 11, 19), composite appeal combinations (Cluster 17), and fear-based messages with gain or loss framing (Clusters 10, 12). The defining feature is direct persuasive prompting without evidential support.

**(4) Contextual Expression Pattern**

This pattern includes posts in which most SAFE elements are weak or absent (approximately 177,000 posts).

Clusters 8 and 18 reflect low-intensity expressions with minimal structure, while Cluster 2 includes weak gain framing or authority cues without clear supporting elements. Overall, this combination reflects low informational density.

**(5) Cluster Validation**

Cosine similarity analysis showed that within-pattern similarity exceeded between-pattern similarity (Figure 4; Table 3). The global average similarity was 0.43 (SD = 0.18), while within-pattern similarity ranged from 0.60 to 0.66 across the four patterns. Similarity was also high within subgroups (e.g., Clusters 0 and 15: 0.90; Clusters 1 and 9: 0.90), supporting the robustness of the grouping.

### *Synergistic and Trade-off Effects of Element Combinations Within Message Patterns*

To examine how message patterns relate to communication effects, we defined a baseline condition for each pattern based on its dominant SAFE structure, typically characterized by two core elements that anchor its primary persuasive logic. Specifically, the Institutional Authority pattern emphasized authoritative sources and expert evidence; the Narrative pattern centered on narrative evidence without explicit sources; the Assertive Appeal pattern lacked both source and evidence; and the Contextual Expression pattern lacked clear appeal and evidence cues.

Within each baseline, we evaluated whether the presence of an additional element combination (S) was associated with differences in communication effects. In conceptual terms, these additional elements are treated as peripheral elements because they extend, rather than define, the core persuasive logic of the baseline pattern. Posts were divided into those satisfying the baseline with S and those satisfying the baseline without S. Communication effect indicators, including likes, comments, and shares, were log-transformed using $\log(x + 1)$, and $\Delta E$ was defined as the difference in mean log-transformed communication effect indicators between the two groups. To estimate statistical uncertainty, we used bootstrap resampling (n = 500) to compute 95% confidence intervals, which is well suited for skewed communication effect indicators (Efron & Tibshirani, 1994; Salganik, 2017). To reduce instability across multiple combination tests, we reported only combinations with N > 300 and bootstrap 95% confidence intervals that did not cross zero. This threshold reduces the influence of small cell sizes, while the confidence interval criterion helps ensure that the reported positive or negative associations are statistically stable. Figure 5 presents representative combinations using a Venn-style structure, highlighting both synergistic and trade-off associations.

**(1) Institutional Authority Pattern**

Combinations introducing metaphor, utilitarian appeal, narrative evidence, or value appeal were positively associated with communication effects (Table 5). For likes, Metaphor + Utilitarian + NarrEv + Value ($\Delta E = 1.097$) and Metaphor + Utilitarian ($\Delta E = 0.976$) were strongest. For comments, Humor + NoFrm ($\Delta E = 1.044$) and Humor + Gain ($\Delta E = 0.770$) performed best. For shares, Metaphor + Utilitarian + NarrEv + Value ($\Delta E = 1.441$) showed the largest increase. Combinations lacking appeal and framing (NoApp + NoFrm) were consistently negative.

**(2) Narrative Pattern**

Emotional and contrastive appeals were positively associated with communication effects (Table 5). For likes, Dual ($\Delta E = 0.587$) and Humor ($\Delta E = 0.572$) were strongest. For comments, humor ($\Delta E = 0.842$) and Humor + Gain ($\Delta E = 0.810$) performed best. For shares, Metaphor + Value + Gain ($\Delta E = 0.572$) was most effective. Adding expert evidence (e.g., ExpEv + Gain) was negatively associated with communication effects.

**(3) Assertive Appeal Pattern**

Strengthening appeals generally increased effects (Table 5). For likes, Dual + Value ($\Delta E = 0.646$) and Dual ($\Delta E = 0.622$) were strongest. For comments, humor ($\Delta E = 0.691$) and Humor + Gain ($\Delta E = 0.682$) performed best. For shares, Value + Utilitarian ($\Delta E = 0.299$) showed a modest positive association with communication effects.

**(4) Contextual Expression Pattern**

This pattern showed weaker communication effects overall (Table 5). For likes, No Source ($\Delta E = 0.438$) and No Source + NoFrm ($\Delta E = 0.430$) showed small increases. Similar patterns appeared for comments. For shares, Gain + No Source ($\Delta E = 0.160$) showed a minor positive effect, while official media cues were negatively associated.

Overall, the effects of additional elements vary across patterns, and the same element may produce positive, weak, or negative associations depending on how it is combined with other elements.

Beyond the effects of specific element combinations, we further examined how the number of additional elements relates to communication effects. Each pattern is anchored in its own baseline structure, and the plotted values represent $\Delta E$, defined as the difference in mean log-transformed effects between posts with peripheral elements and those with only the baseline. The x-axis indicates the number of peripheral elements added beyond the baseline structure, where each baseline consists of two core elements (Figure 6).

Based on statistically significant combinations, we aggregated ΔE by the number of added elements (k) and compared average communication effects across patterns and indicators (Figure 6). The results reveal a clear non-linear relationship between k and effects. ΔE increases substantially as k rises from low levels, particularly when two to four peripheral elements are added beyond the baseline, indicating that effective message design depends not only on core structures but also on the incorporation of a moderate number of peripheral elements. This pattern is further supported by the top-performing combinations (Table 5), where higher ΔE values are also concentrated within this range, suggesting diminishing returns beyond it.

However, this upward trend does not continue proportionally. Beyond four elements, the peripheral increase in ΔE becomes notably smaller, and in some cases stabilizes. This suggests that their incremental benefit diminishes as complexity increases.

*Variation in Effects Across Source Influence Levels*

To examine how source influence moderates the effects of message combinations, we grouped accounts into follower-based tiers and identified the combinations most strongly associated with effects within each tier (Table 6).

Across the Institutional Authority pattern, high-visibility accounts tended to achieve stronger effects with relatively simple combinations, such as Dual + NoFrm for likes and Sex for comments. By contrast, low-visibility accounts showed their strongest associations for more elaborate combinations, such as Metaphor + Utilitarian + NarrEv + Gain for both likes and shares. Middle-visibility accounts generally fell between these two patterns.

A similar tendency appeared in the Narrative and Assertive Appeal patterns. In high-visibility accounts, relatively compact combinations such as Humor, Sex, or Dual were often associated with stronger effects. In low-visibility accounts, more complex combinations that integrated multiple elements, such as Utilitarian + Value + Gain or Comparative + Gain, were more prominent among the strongest-performing combinations.

The Contextual Expression pattern showed weaker differentiation across visibility groups,

with only modest effects differences overall.

To summarize this pattern more formally, we compared the number of elements in statistically significant combinations across source influence groups (Table 6). Kruskal–Wallis tests indicated significant variation in combination complexity, H = 15.92, p < .001. Dunn's post-hoc tests showed that bottom-tier accounts involved more complex significant combinations than top-tier accounts (p < .001), and middle-tier accounts also differed from top-tier accounts (p < .05), whereas the difference between middle- and bottom-tier accounts was not significant.

Consistent with this pattern, bottom-tier accounts exhibited higher average complexity (M = 2.03, SD = 0.79), compared to middle-tier (M = 1.86, SD = 0.62) and top-tier accounts (M = 1.67, SD = 0.59). High-visibility accounts typically relied on one to two additional elements (e.g., Humor, Dual), whereas low-visibility accounts more frequently involved two to four elements (e.g., Metaphor + Utilitarian + NarrEv + Gain).

## Discussion
### *Persuasion Is Organized Around Distinct Multi-Element Baseline Patterns*

In response to RQ1, four recurring patterns were identified: Institutional Authority Pattern (IAP), Narrative Pattern (NP), Assertive Appeal Pattern (AAP), and Contextual Expression Pattern (CEP). These patterns represent distinct SAFE baselines that organize persuasion in different ways. IAP is credibility-centered, NP is narrative-centered, AAP is appeal-driven without explicit source or evidence cues, and CEP reflects low-intensity expression with minimal persuasive structure. Together, these findings suggest that persuasion in social media is organized around structured baseline patterns rather than isolated elements, and that communication and persuasive effects depend on how peripheral elements are combined around these baselines.

First, synergy occurs when peripheral elements reinforce baseline logic. In IAP, combinations integrating narrative, metaphor, and value or utilitarian cues produce the strongest communication effects, suggesting that authoritative credibility is more effective when paired with interpretable and relatable elements rather than presented in purely formal terms. This is consistent with research showing that expert credibility alone is often insufficient unless supported by accessible, audience-oriented design (Petty & Cacioppo, 1986; De Meulenaer et al., 2017; Limbu & McKinley, 2025). By contrast, combinations lacking appeals or frames weaken effectiveness, indicating that credibility alone may not stimulate audience response. This is especially evident for sharing, where combinations that enhance interpretability and perceived usefulness are associated with stronger diffusion.

Second, trade-offs emerge when peripheral elements conflict with baseline logic. In NP, emotional elements such as humor and metaphor are positively associated with communication effects, consistent with the view that narrative persuasion works through immersion, resonance, and affective involvement (Green & Brock, 2000; Dahlstrom, 2014; Bilandzic et al., 2020). However, introducing expert evidence tends to reduce effectiveness, suggesting that analytical cues may interrupt rather than strengthen narrative processing. The negative effect therefore reflects incompatibility with baseline logic rather than simply adding more information. At the same time, for sharing, combinations integrating metaphor and value framing perform better, indicating that narrative persuasion can benefit from added structure when it remains affectively and interpretively coherent (Braddock & Dillard, 2016; Hornikx, 2018).

Third, the effects of peripheral elements vary across effects indicators. In AAP, emotional elements mainly amplify immediate responses such as likes and comments but are less effective for sharing, whereas value- and utility-oriented combinations are relatively more conducive to diffusion. This suggests that affective prompting is effective for generating reaction, but less sufficient for motivating retransmission unless combined with elements that increase perceived usefulness or relevance, a distinction also noted in research on emotional and informational message effects in social media health communication (Dillard & Nabi, 2006; Samuel, Kuijpers & Bleakley, 2024; Kuo & Chen, 2025). By contrast, CEP shows consistently limited responsiveness, suggesting that low-intensity baselines constrain the contribution of peripheral elements overall. Rather than functioning as a persuasion-oriented pattern, CEP appears to operate as a form of ambient communication whose primary role is routine information maintenance rather than active persuasion. In this context, explicit source or evidence cues are not necessarily advantageous and may even reduce communicative efficiency, indicating that minimal combinations are better aligned with its functional role.

***Source Influence Conditions the Complexity and Alignment of Persuasive Combinations***

The results for RQ3 indicate that the relationship between message patterns and communication effects is conditioned by source influence, operationalized as follower-based prominence. As in RQ2, each pattern remains anchored in a dominant SAFE baseline, but source influence shapes how additional elements align with this baseline to generate stronger communication effects.

First, source influence alters the level of complexity required for effective alignment. In Institutional Authority Persuasion (IAP), high-influence accounts show stronger communication effects using relatively simple extensions (e.g., Dual + NoFrm; Dual + NarrEv), whereas lower-influence accounts require more elaborated combinations (e.g., Metaphor + Utilitarian + NarrEv + Gain). This suggests that when credibility is externally supported by source prominence, fewer additional elements are needed, whereas in its absence, elements must collectively compensate.

Second, alignment strategies vary across influence tiers. In Narrative Persuasion (NP), high-influence accounts perform strongly with minimal emotional cues (e.g., Humor, Sex), whereas lower-influence accounts rely on combinations that introduce additional structure (e.g., Comparative + Gain; Utilitarian + Value + Gain). This indicates that emotional resonance can sustain communication effects when supported by source prominence, while lower-influence accounts require additional framing to stabilize interpretation.

Third, indicator-specific divergence appears across influence tiers, particularly for sharing. In AAP, high-influence accounts show stronger communication effects with relatively simple affective combinations (e.g., Metaphor + Value), whereas lower-influence accounts rely more on utilitarian and value-oriented elements (e.g., Value + Utilitarian + Gain). This mirrors RQ2, where likes and comments are affect-driven, while sharing depends on perceived utility.

Finally, CEP shows limited variation across influence tiers. Even among high-influence accounts, only modest gains are observed, and stronger persuasive elements do not substantially improve effects. This suggests that the low-intensity baseline constrains integration regardless of source influence.

### *Implications for Theory and Practice*

Theoretically, this study extends prior research on health message effects by shifting attention from isolated message elements to structured combinations. The findings suggest that source cues, appeals, frames, and evidence do not operate as independent components, but are embedded in broader message patterns that condition their association with communication and persuasive effects. To organize these findings, this study proposes a Core–Periphery–Environment framework (Figure 7). In this framework, core structures refer to stable baseline combinations, peripheral elements modify the association between these baselines and communication effects, and environmental conditions, such as source influence, shape the level of complexity associated with audience response. This framework should be understood as a middle-range analytical framework rather than a causal model. It highlights that communication effects are not simply improved by adding more elements. Instead, effective combinations depend on whether peripheral elements align with the core logic of a message. When added elements reinforce the baseline structure, they are more likely to be associated with stronger communication effects; when they conflict with that structure, trade-offs may occur. The findings also suggest a bounded role of complexity. Moderate complexity, especially combinations involving two to four peripheral elements, is more often associated with advantages in communication effects, whereas excessive complexity shows diminishing returns.

From a practical perspective, stronger health communication effects depend on matching communication goals with appropriate baseline patterns and aligning peripheral elements with their underlying logic. Different goals call for distinct message patterns and element combinations. Authority-based messages tend to show stronger communication effects when credibility cues are combined with value or utilitarian elements, whereas narrative-based messages benefit from emotional coherence and may be weakened by analytical cues. In contrast, the Contextual Expression Pattern (CEP), which corresponds to low-intensity informational goals, is better served by simpler combinations, where limited source or evidence cues preserve clarity and efficiency. Increasing the number of elements does not guarantee stronger communication effects and may weaken them when poorly matched. Strategies should also adapt to source influence. High-visibility accounts can generate strong communication effects with relatively simple combinations, whereas lower-visibility accounts benefit from more elaborated combinations, indicating that complexity can function as a compensatory strategy when source influence is limited. Affective combinations are more strongly associated with likes and comments, whereas sharing is more strongly associated with combinations that enhance interpretability and perceived utility. The findings also inform AI-assisted content generation. As AI systems increasingly rely on template- and prompt-based structures (Laba, 2024), incorporating structured element combinations can support context-sensitive design that selects baseline patterns, aligns elements, and adjusts complexity based on source influence.

### *Strengths and Limitations*

This study provides a large-scale analysis of social media health communication and shows how persuasive elements are organized into recurring combinations. By combining theory-guided coding, clustering, and communication effects comparison, it offers a systematic account of multi-element message structures in a real-world platform environment. Several limitations remain. First, the study is observational, so ΔE should be interpreted as differences in communication effects associated with element combinations rather than causal effects. Second, likes, comments, and shares capture communication effects, but they do not directly measure attitude or behavior change. Third, the analysis focuses on textual content, while visual and video cues may also shape engagement. Finally, source influence is measured by follower-based visibility, which reflects platform prominence but not necessarily perceived credibility or trust. Future research could address these limitations through experiments, surveys, and multimodal analysis.

## Funding

This work was supported by the General Project of the Ministry of Education Foundation on Humanities and Social Sciences (Grant No. 23YJA860011), the Fundamental Research Funds for the Central Universities (Grant No. 1243200012), and the Guangdong Philosophy and Social Science Foundation Regular Project (Grant No. GD24XXW02).

## Declaration of Interest

The authors report no conflicts of interest.

Table 1. SAFE Framework Encoding and Prompt Concepts

| Dimension | Code | Abbreviation | Concept |
|---|---|---|---|
| **Source** | Expert | Expert Source(Exp) | Results or analyses from recognized professionals, scholars, or authoritative institutions (Vraga & Bode, 2017). |
| | Official Media | Official Media(OffM) | Information published by government, public health agencies, or credible news media (Vraga & Bode, 2017). |
| | No Source | No Source(NoSrc) | Lack of clear or identifiable source support. |
| **Appeal** | Sex Appeal | Sex | Emotional appeal based on attraction, suggestiveness, or sensual cues (Dahlen et al., 2010). |
| | Fear Appeal | Fear | Emphasizing potential negative consequences to induce fear or anxiety (Witte, 1992). |
| | Humor Appeal | Humor | Using humor or ironic contrast to create positive emotional responses (Nabi, 2003; Dillard & Nabi, 2006). |
| | Value Appeal | Value(Valu) | Promoting health-related norms, responsibility, or prosocial values (Hornik et al., 2016). |
| | Utilitarian Appeal | Utilitarian(Util) | Emphasizing functional benefits, efficacy, or practical utility (Johar & Sirgy, 1991; Taylor & Knibb, 2013). |
| | Dual Appeal | Dual | Presenting both supporting and opposing viewpoints (Rucker et al., 2008). |
| | Comparative Appeal | Comparative(Comp) | Comparing alternatives or highlighting relative advantages (Dahlen et al., 2010). |
| | Metaphor Appeal | Metaphor(Met) | Using symbolic or figurative expressions to convey abstract concepts (Limbu & McKinley, 2025). |
| | No Appeal | NoApp | No clearly identifiable appeal. |
| **Frame** | Gain Frame | Gain | Highlighting benefits of adopting recommended behavior (Rothman & Salovey, 1997). |
| | Loss Frame | Loss | Emphasizing risks of not adopting recommended behavior (Rothman & Salovey, 1997). |
| | No Frame | NoFrm | No clear framing strategy. |
| **Evidence** | Narrative Evidence | NarrEv | Using stories or personal experiences (Allen & Preiss, 1997; Zebregs et al., 2015). |
| | Statistical Evidence | StatEv | Presenting numerical or aggregated data (Zillmann & Brosius, 2000). |
| | Expert Evidence | ExpEv | Referencing expert opinions or scientific research (Hoeken & Hustinx, 2009). |
| | Causal Evidence | CausEv | Describing cause–effect relationships (Hoeken & Hustinx, 2009). |
| | No Evidence | NoEv | No clear evidence provided. Single choice only. |

Table 2. Cluster Aggregation Systematized Persuasion Model Definitions

| Primary | Secondary | Basic Cluster | Representative Combination | Number |
|---|---|---|---|---|
| **Institutional Authority Persuasion (IAP)** | Expert endorsement type | Expert value endorsement (Cluster 16) | Exp + Valu + Gain + ExpEv | 119,364 |
| | | Expert utilitarian endorsement (Cluster 15) | Exp + Util + Gain + ExpEv | 37,571 |
| | | Expert composite appeal (Cluster 0) | Exp + Valu×Util + Gain + ExpEv | 34,429 |
| | Neutral representation type | Expert direct endorsement (Cluster 7) | Exp + NoApp + Gain + ExpEv | 39,940 |
| | | Expert neutral representation (Cluster 14) | Exp + NoApp + NoFrm + ExpEv | 29,429 |
| | Official declaration type | Official value declaration (Cluster 13) | OffM + (Valu/Util) + Gain + (NoEv/ExpEv) | 48,295 |
| | | Authority convergence (Cluster 20) | OffM×Exp + Valu + Gain + (Stat/ExpEv) | 65,991 |
| **Narrative Persuasion (NP)** | Practical narrative type | Value narrative (Cluster 3) | NoSrc + Valu + Gain + Narr | 73,268 |
| | | Utilitarian narrative (Cluster 9) | NoSrc + Util + Gain + Narr | 25,033 |
| | | Composite narrative appeal (Cluster 1) | NoSrc + Valu×Util + Gain + Narr | 41,380 |
| | Everyday narrative type | Everyday gain narrative (Cluster 4) | NoSrc + NoApp + Gain + Narr | 22,585 |
| | | Neutral narrative (Cluster 5) | NoSrc + NoFrm + Narr | 19,193 |
| | Authority narrative type | Authority narrative hybrid (Cluster 6) | Exp + Valu + Gain + Narr | 20,416 |
| **Assertive Appeal (AA)** | Utilitarian assertive type | Value assertive (Cluster 19) | NoSrc + Valu + Gain + NoEv | 284,927 |
| | | Utilitarian assertive (Cluster 11) | NoSrc + Util + Gain + NoEv | 88,402 |
| | Practical composite type | Value–utilitarian dual assertive (Cluster 17) | NoSrc + Valu×Util + Gain + NoEv | 68,448 |
| | Fear assertive type | Fear mobilization (Cluster 10) | NoSrc + Fear + Gain + NoEv | 21,514 |
| | | Fear deterrence (Cluster 12) | NoSrc + Fear + Loss + NoEv | 36,975 |
| **Contextual Expression (CE)** | Leisure context type | Absent expression (Cluster 18) | NoSrc + NoApp + NoFrm + NoEv | 153,028 |
| | | Positive expression (Cluster 8) | NoSrc + NoApp + Gain + NoEv | 81,477 |
| | Fuzzy positive type | Fuzzy positive (Cluster 2) | (OffM/Exp) + NoApp + Gain + NoEv | 25,012 |

Table 3. Post-Hoc Cluster Average Cosine Similarity Verification

| Model | Secondary | Relative Cluster Correlation | Average | GAS |
|---|---|---|---|---|
| IAP | Expert Endorsement Type | 0–15 (0.90), 0–16 (0.89), 15–16 (0.75) | 0.60 | 0.43 (SD = 0.18) |
|  | Neutral Representation Type | 7–14 (0.75) |  |  |
|  | Official Declaration Type | 13–20 (0.65) |  |  |
| NP | Practical Narrative Type | 3–9 (0.75), 1–9 (0.90), 1–3 (0.86) | 0.65 |  |
|  | Everyday Narrative Type | 4–5 (0.75) |  |  |
|  | Authority Narrative Type | 6 (single cluster) |  |  |
| AA | Practical Assertive Type | 11–19 (0.75) | 0.65 |  |
|  | Practical Composite Type | 17 (single cluster) |  |  |
|  | Fear Assertive Type | 10–12 (0.64) |  |  |
| CE | Leisure Context Type | 8–18 (0.75) | 0.66 |  |
|  | Fuzzy Positive Type | 2 (single cluster) |  |  |

Table 4. Optimal and Least Optimal Persuasion Element Combinations in Models

| Model | Polarity | Likes | Comments | Shares |
|---|---|---|---|---|
| IAP | Positive | Met+Util+Narr+Valu (1.097); Met+Util+Narr+Gain (1.094) | Hum+NoFrm (1.044); Hum+Gain (0.770) | Met+Util+Narr+Valu (1.441); Met+Narr (1.056) |
| IAP | Negative | NoApp+NoFrm (−0.307); NoApp (−0.301) | NoApp+NoFrm (−0.172); NoApp (−0.170) | NoApp+NoFrm (−0.425); NoApp (−0.417) |
| NP | Positive | Dual (0.587); Hum (0.572) | Hum (0.842); Hum+Gain (0.810) | Met+Valu+Gain (0.572); Valu+Met+Gain (0.572) |
| NP | Negative | ExpEv+Gain (−0.257); ExpEv+NoFrm (−0.246) | ExpEv+Gain (−0.543); ExpEv+NoFrm (−0.512) | NoApp+NoFrm (−0.483); NoApp (−0.455) |
| AA | Positive | Dual+Valu (0.646); Dual (0.622) | Hum (0.691); Hum+Gain (0.682) | Valu+Util (0.299); Valu+Util+Gain (0.299) |
| AA | Negative | NoApp (−0.200); NoApp+Gain (−0.188) | Gain (−0.313); Valu+Util (−0.307) | Hum (−0.256); NoApp (−0.253) |
| CE | Positive | NoSrc (0.438); NoSrc+NoFrm (0.430) | NoSrc (0.263); NoSrc+NoFrm (0.259) | Gain+NoSrc (0.160); Gain (0.158) |
| CE | Negative | OffM+Gain (−0.553); OffM (−0.550) | OffM+Gain (−0.358); OffM (−0.352) | OffM (−0.153); OffM+Gain (−0.148) |

Table 5. Optimal Combinations of Source Influence on Persuasion Modes

| Model | Source | Likes | Comments | Shares |
|---|---|---|---|---|
| IAP | Top | Dual + NoFrm (0.82); Dual + Narr (0.81) | Sex (1.14); Sex + Gain (0.80) | Dual + Narr (0.57); Valu (0.56) |
| | Middle | Met + Util + Valu + Gain (1.29); Met + Util + Valu (1.26) | Sex + Util (0.91); Hum + Gain + Valu (0.82) | Dual + Caus (0.74); Comp + Dual (0.46) |
| | Bottom | Met + Util + Narr + Gain (1.34); Met + Util + Narr (1.33) | Hum + NoFrm (0.98); Narr + Hum (0.90) | Met + Util + Narr + Gain (1.64); Met + Util + Narr (1.62) |
| NP | Top | Hum (1.17); Sex (1.12) | Sex (1.24); Hum (1.16) | Dual (0.50) |
| | Middle | Met + Gain + Valu (0.65); Met + Valu (0.61) | Util (0.07); Gain + Util (0.07) | Valu + Met + Gain (0.86); Valu + Met (0.79) |
| | Bottom | Comp (0.41); Comp + Gain (0.40) | Comp + Gain (0.49) | Util + Valu + Gain (0.63); Valu + Util (0.63) |
| AA | Top | Met + Valu (0.83); Met + Valu + Gain (0.83) | Sex (1.48); Sex + Gain (1.38) | Met + Valu + Gain (0.88); Met + Valu (0.85) |
| | Middle | Dual + Valu (0.62); Dual (0.61) | Sex + Util (0.76); Dual (0.61) | Gain + Met (0.53); Met (0.46) |
| | Bottom | Hum + Valu (0.43); Comp + Gain + Util (0.30) | Loss (0.21); Loss + Fear (0.19) | Valu + Util + Gain (0.47); Valu + Util (0.47) |
| CE | Top | NoFrm (0.34) | Exp (0.60); Gain + Exp (0.42) | Loss (0.37); Gain (0.13) |
| | Middle | Exp (0.59); Gain + Exp (0.52) | – | Gain + Exp (0.34) |
| | Bottom | – | Loss (0.19) | Exp (0.12) |

Table 6. Combination Quantity Statistics

| Level | Count | Avg. (Comb.) | SD | Min | Max | Dunn's p |
|---|---|---|---|---|---|---|
| **Mid** | 154 | 1.86 | 0.62 | 1 | 4 | Mid–Head: 0.03; Mid–Tail: 0.38 |
| **Head** | 126 | 1.67 | 0.59 | 1 | 3 | Head–Tail: 0.000 |
| **Tail** | 155 | 2.03 | 0.79 | 1 | 4 | — |

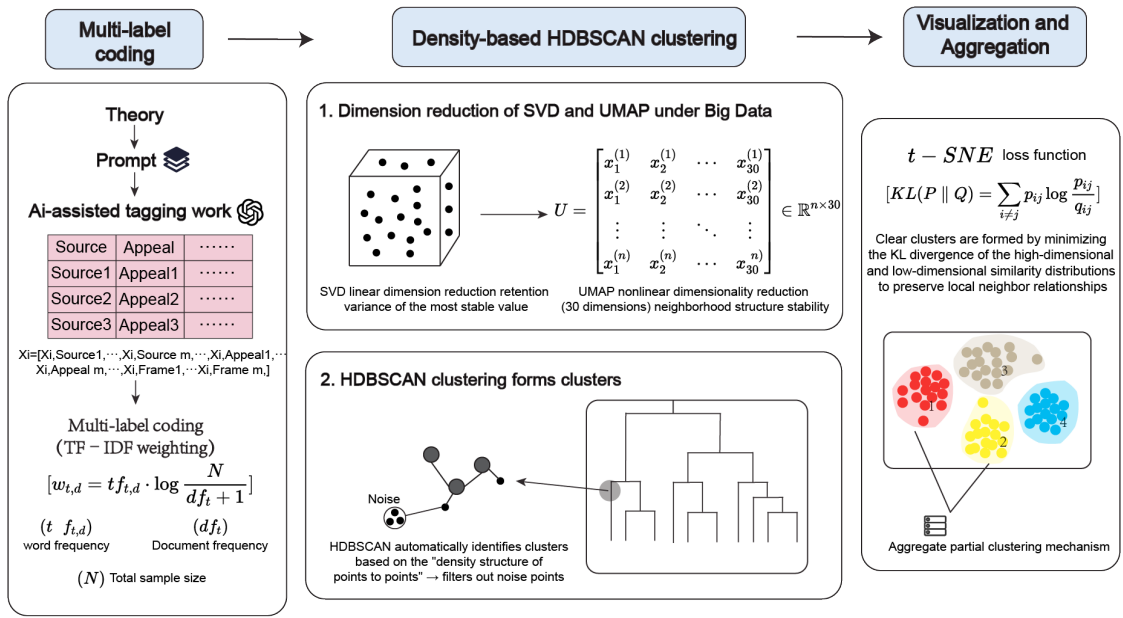

Figure 1. Theory-guided Multi-Element Clustering Computational Pathway

*Note.* Texts are first encoded using SAFE-based theoretical features and embedded into a theoretical space, followed by HDBSCAN-based density clustering and abductive aggregation into persuasion mechanisms.

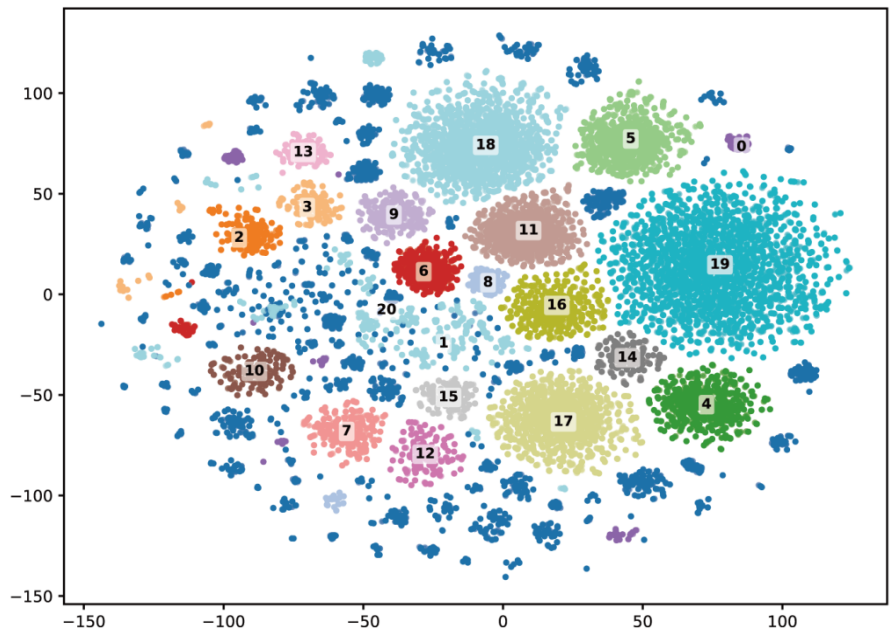

Figure 2. HDBSCAN-based Density Clustering Map

*Note.* Each point represents a post embedded in the SAFE theoretical space. Colors indicate cluster membership identified by HDBSCAN; gray points denote noise.

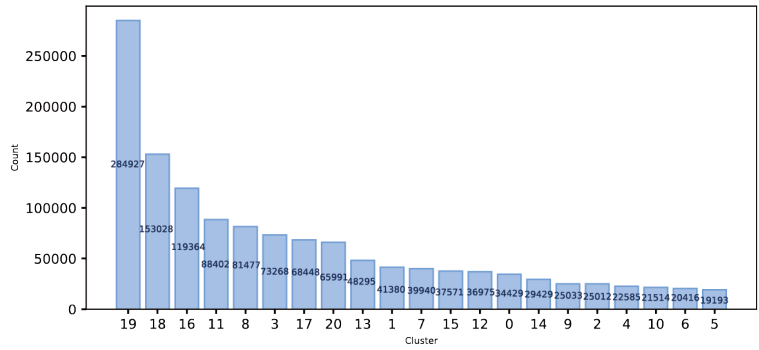

Figure 3. Cluster Quantity Statistical Map

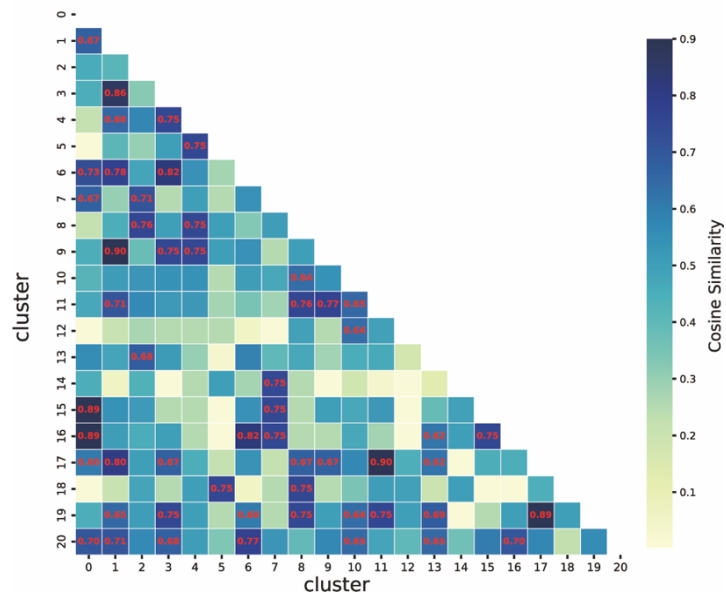

Figure 4. SAFE Cluster Cosine Similarity Test Heat Map (CS>0.6)

*Note.* Values represent average cosine similarity between clusters based on SAFE dimensions. Higher values indicate stronger internal coherence.

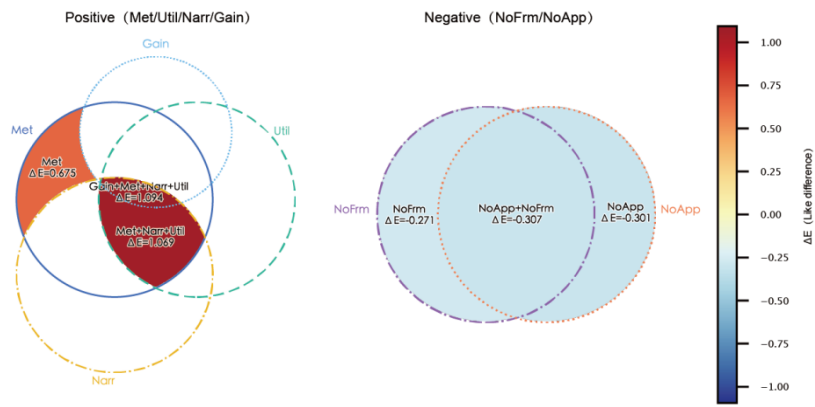

Figure 5. Example Visualization of Venn-style Significant Combination Overlap

*Note.* Red areas indicate reinforcing effects (positive ΔE), blue areas indicate inhibiting effects (negative ΔE). Only combinations with N > 300 are displayed.

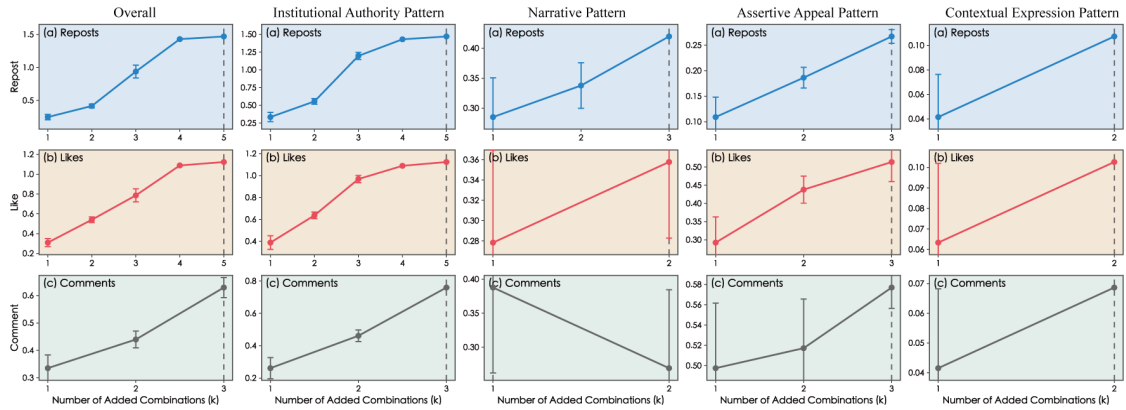

Figure 6. Relationship Between the Number of Additional Elements and Communication Effects Relative to Pattern-Specific Baselines

*Note.* ΔE denotes the difference in mean log(x + 1) communication effect indicators between posts with additional elements and those containing only the pattern-specific baseline. The x-axis represents the number of additional elements (k); only statistically significant combinations are included, and error bars indicate bootstrap standard errors (n = 500).

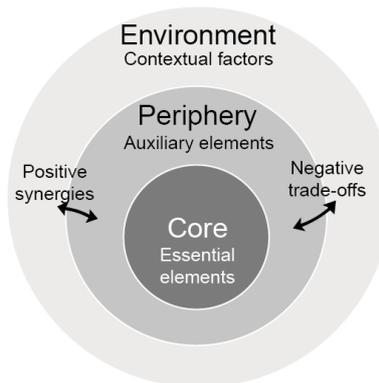

Figure 7. Systems Persuasion

*Note.* Core mechanisms represent stable persuasion structures; peripheral elements modulate effectiveness; environmental factors condition feasible combinations.